\documentclass[a4paper,11pt]{article}
\usepackage{jinstpub} %
\usepackage{lineno}

\usepackage[
    separate-uncertainty=true, 
    round-mode=uncertainty,
    round-precision=1,
]{siunitx}

\title{\boldmath OmniJet-$\alpha_C$: Learning point cloud calorimeter simulations using generative transformers}

\author[a]{Joschka Birk}
\affiliation[a]{
	Institute for Experimental Physics, Universität Hamburg \\
	Luruper Chaussee 149, 22761 Hamburg, Germany
}

\author[b]{Frank Gaede}
\affiliation[b]{
    Deutsches Elektronen-Synchrotron DESY, \\
    Notkestr. 85, 22607 Hamburg, Germany
}

\author[a,1]{Anna Hallin\note{Corresponding author.}}
\emailAdd{anna.hallin@uni-hamburg.de}

\author[a]{Gregor~Kasieczka}

\author[a]{Martina Mozzanica}

\author[a]{Henning Rose}

\abstract{We show the first use of generative transformers for generating calorimeter showers as point clouds in a high-granularity calorimeter.
    Using the tokenizer and
    generative part of the \textsc{OmniJet}-$\alpha$ model, we represent the hits in the detector as
    sequences of integers. This model allows variable-length sequences, which
    means that it supports realistic shower development and does not need to be
    conditioned on the number of hits. Since the tokenization represents the
    showers as point clouds, the model learns the geometry of the showers
    without being restricted to any particular voxel grid. }


\arxivnumber{2501.05534} %

\begin{document}
\maketitle
\flushbottom

\section{Introduction}

Machine learning (ML) methods have been a common ingredient in particle physics
research for a long time, with neural networks being applied to object
identification already in analyses at LEP~\cite{Behnke:1995vc}.
Since then, the range of applications has grown drastically, with ML methods
being developed and used for example in tagging~\cite{Draguet:2912358,Karwowska:2024xqy,Mondal:2024nsa}, anomaly detection~\cite{ATLAS:2020iwa,ATLAS:2023azi,ATLAS:2023ixc,CMS:2024nsz}, 
individual reconstruction stages like particle 
tracking~\cite{Burleson:2882507,ATL-PHYS-PUB-2024-018,Correia:2024kal} or even full event interpretation and reconstruction~\cite{GarciaPardinas:2023pmx}.
Another important use case for ML in high energy physics (HEP) is detector
simulation. With the increasing luminosity of the large-scale experiments in
HEP, the computational cost of high-precision Monte-Carlo (MC) simulations is
going to exceed the available computing resources~\cite{Adelmann:2022ozp}.
Generative methods have the potential to significantly reduce this resource
requirement, which is why a considerable amount of research has been spent on
exploring machine learning architectures for detector simulation \cite{Butter:2022rso,
Krause:2024avx}. Examples include GANs
\cite{Paganini:2017hrr, Paganini:2017dwg, deOliveira:2017rwa, Erdmann:2018kuh,
Musella:2018rdi, Erdmann:2018jxd, Belayneh:2019vyx, Butter:2020qhk, Javurkova:2021kms,
Bieringer:2022cbs, Hashemi:2023ruu, 2024_atlas}, 
variational autoencoders (VAEs) and their variants
\cite{Buhmann:2020pmy, Buhmann:2021lxj, Buhmann:2021caf, 2024_atlas,Cresswell:2022tof,Diefenbacher:2023prl},
normalizing flows and various
types of diffusion \mbox{models ~\cite{sohldickstein2015deep,
song2020generative_estimatingGradients,
song2020improved_technieques_for_sorebased_geneerative, ho2020denoising,
song2021scorebased_generativemodelling, Mikuni:2022xry_CaloScore, Buhmann:2023bwk,
Acosta:2023zik, Mikuni:2023tqg,
Amram:2023onf, Chen:2021gdz, Krause:2021ilc, 
Krause:2021wez, schnake2022_pointFlow, Krause:2022jna,
Diefenbacher:2023prl, 
Xu:2023xdc, Buckley:2023daw,Ernst:2023qvn, OmanaKuttan:2024mwr,Favaro:2024rle, Brehmer:2024yqw}}. 

Most ML methods in HEP are designed, developed and trained for
very specific tasks. The focus on specialized models means that the full potential of the
vast datasets we have access to is not being utilized. Furthermore, while these
models may be more resource efficient than the traditional methods they seek to
enhance or replace, developing and training each model from scratch still
requires significant amounts of both human and computational resources.
For reasons like these, there has been a growing interest in developing
foundation models for particle
physics~\cite{Kishimoto:2023cys,Qu:2022mxj,Golling:2024abg,Birk:2024knn,Harris:2024sra,Mikuni:2024qsr,Wildridge:2024yeg,Amram:2024fjg,Ho:2024qyf} 
in the past couple of years. A foundation model is a machine learning model that has been pre-trained on a
large amount of data, and can then be fine-tuned for different downstream
tasks~\cite{bommasani2022opportunities}. The idea behind utilizing pre-trained models is that their outputs can
significantly enhance the performance of downstream tasks, yielding better
results than if the model were to be trained from scratch. While the models
mentioned above have focused on exploring different tasks in specific
subdomains, like jet physics, a more ambitious goal eventually would be to
develop a foundation model for all tasks in all subdomains, including for
example tracking, shower generation and anomaly detection in general (not
restricted to jets). The hope would be that it could then utilize the full
amount of diverse data from our experiments, to boost the performance of all
possible downstream tasks. The first step towards such a model must be to be
able to handle tasks from different subdomains in the same computational
framework. 

In this work, we apply the generative
part of \textsc{OmniJet}-$\alpha$~\cite{Birk:2024knn}, originally developed for jet physics, to a
completely different subdomain: electromagnetic shower generation in collider
calorimeters. We show that the \textsc{OmniJet}-$\alpha$ architecture and workflow also works for
generating showers, opening up the possibility of exploring transfer learning
for showers in a setting that has already proved successful in the context of
jet physics. This is the first example of an autoregressive generative model
utilizing the GPT architecture for calorimeter point clouds (as opposed to the
fixed calorimeter geometries of Ref.~\cite{Liu:2024kvv}).
We denote this extended model capable of handling showers as \textsc{OmniJet}-$\alpha_C$ (\textsc{OmniJet}-$\alpha$
Calorimeter). Showing that we can use the same framework for two very different
subdomains is an important step towards developing a foundation model for all
computing and data analysis tasks in particle physics.

This paper is organized as follows. Section~\ref{sec:dataset} describes the
dataset used, section~\ref{sec:methods} the experimental setup, and
section~\ref{sec:results} presents the results. Finally, we offer our
conclusions in section~\ref{sec:conclusions}.

\section{Dataset}
\label{sec:dataset}
The International Large Detector (ILD) \cite{ILDConceptGroup:2020sfq} is one of two detector 
concepts proposed for the International Linear Collider (ILC)
\cite{Behnke:2013xla}, an electron-positron collider that is initially operated
at $250$~GeV center-of-mass energy and extendable to higher energies up to
$1$~TeV.
ILD is optimized for the Particle Flow Algorithm~\cite{Thomson:2011zz} that aims
at reconstructing every individual particle. The detector therefore combines
precise tracking and vertexing capabilities with good hermiticity, and highly
granular sandwich calorimeters. The electromagnetic calorimeter of ILD (the
\mbox{Si-W}~ECAL \cite{Suehara:2018mqk}) consists of 20 layers with 2.1~mm thick
W-absorbers followed by 10 layers with 4.2~mm W-absorbers, all interleaved with
0.5~mm thick Si-sensors that are subdivided into 5~mm~$\times$~5~mm cells.

The dataset used in this work was originally created for
Ref.~\cite{Buhmann:2020pmy}, where more details on the detector and simulation
can be found. While the dataset itself is not publicly available, 
it can be fully recreated by following the simulation and processing instructions provided in Ref.~\cite{Buhmann:2020pmy}. 
Showers of photons with initial energies uniformly distributed
between $10-100$~GeV are simulated with Geant4~\cite{Agostinelli:2002hh} using a
detailed and realistic detector model implemented in DD4hep~\cite{Frank:2014zya}.
The resulting showers are projected into a regular 3D grid with 30×30×30 = 27 000 voxels.
The 3D-grid data is converted into a point cloud format, where each point has
four features: the $x$- and $y$-position (transverse to the incident
particle direction), the $z$-position (parallel to the incident particle
direction), and the energy. On average, each shower contains approximately 930, but not more than 1700 points with non-zero energy depositions, representing only a small fraction of the total 27,000 voxels in the grid. 
The incoming photon enters the calorimeter at
perpendicular incident angle from the bottom at $z = 0$ and traverses along the
$z$-axis, hitting cells in the center of the $x$-$y$~plane.
A staggered cell geometry results in small shifts between the layers. 

We preprocess the four input features ($x$, $y$, $z$ and energy) 
by standardization. The energy feature is log-transformed before being scaled
and shifted, which has the additional advantage that generated energies are by
design non-negative.

The dataset has 950\,000 samples, of which 760\,000 are used for training, 95\,000
for validation, and 95\,000 as test samples.

\section{Methods}
\label{sec:methods}
This work uses the workflow of \textsc{OmniJet}-$\alpha$~\cite{Birk:2024knn}, which is a foundation
model originally developed for jet physics. We do not use a pretrained version of \textsc{OmniJet}-$\alpha$, but rather implement the same autoregressive architecture and train it from scratch for generating calorimeter showers.
\textsc{OmniJet}-$\alpha$ uses a
VQ-VAE~\cite{oord2018neural,bao2022beit,Golling:2024abg,huh2023straightening}
to tokenize the input features. The VQ-VAE transforms high-dimensional features 
into discrete latent representations by encoding the data and quantizing it 
to the nearest vectors in a learned codebook.
The constituents of the jets, or in this case
the voxel hits of the showers, are represented as a sequence of integers, which correspond to codebook vectors. A start token and a stop token are added to the beginning and the end of each sequence. These are special tokens that are needed for the autoregressive generation, as described in Section~\ref{sec:generation}. The
sequences are used as input for the generative model, which is a
GPT-style~\cite{Radford2018ImprovingLU} model. Since the model only expects
integers, it is not dependent on a specific type of data as input as long as it
can be represented in this format. Moreover, the model accepts variable-length
sequences, which means that it can be used equally well for jets with a variable
number of constituents as for showers with a variable number of hits.
The training target of the model is
next token prediction, that is, it learns the probability of each token given a
sequence of previous tokens, $p(x_i|x_{i-1},...,x_0)$. This means that it is
straightforward to use the trained model for autoregressive generation, where
each new token is generated conditioned on the previous ones in the sequence.
While \textsc{OmniJet}-$\alpha$ also has classification capabilities,
this work only focuses on the generative part. One key feature of \textsc{OmniJet}-$\alpha$ is that it learns
the sequence length from context. This removes the need for specifying the
number of elements in the sequence beforehand.

The VQ-VAE and generative model were trained using the hyperparameters described in
Appendix~\ref{sec:appendix_hyperparams}. For the VQ-VAE, the best epoch was
selected via lowest validation loss. After training, the VQ-VAE was frozen. The
input data was tokenized using this model, and then fed into the generative model for
training. Here again the epoch with the lowest validation loss was chosen as the
best epoch. New showers in the form of integer sequences were then generated
using this final generative model, and the frozen VQ-VAE was used to decode these
integer sequences back into physical space.

\section{Results}
\label{sec:results}

In the following we will present the results of the training of the VQ-VAE and the generative model.
For comparison we use the test dataset, which the models never saw during training. 
As a benchmark for shower generation the performance of \textsc{OmniJet}-$\alpha_C$ is compared to two state-of-the-art generative networks: one
point cloud model, \textsc{CaloClouds~II}~\cite{Buhmann:2023kdg}, and one
fixed-grid model, \textsc{L2LFlows}~\cite{Buss:2024orz}. \textsc{CaloClouds~II} is
a continuous time score-based diffusion model that has been further distilled into a consistency model (CM), whereas
\textsc{L2LFlows} is a flow-based model using coupling flows with convolutional
layers. \textsc{L2LFlows} has already been trained on this dataset in~\cite{Buss:2024orz}, and showers generated by this model were provided to us directly by the authors. For \textsc{CaloClouds~II} however, no such training was available. Instead we ran this training ourselves, using the same hyperparameters as in~\cite{Buhmann:2023kdg} with the exception of training the diffusion model for 3.5\,M iterations instead of 2\,M, and the consistency model for 2.5\,M iterations instead of 1\,M. This is the first time \textsc{CaloClouds~II} has been trained on a dataset in which the granularity matches the one of the calorimeter. 

\subsection{Token quality}
\label{sec:token_quality}
\begin{figure}[htbp]
   \centering 
   \includegraphics[width=\textwidth]{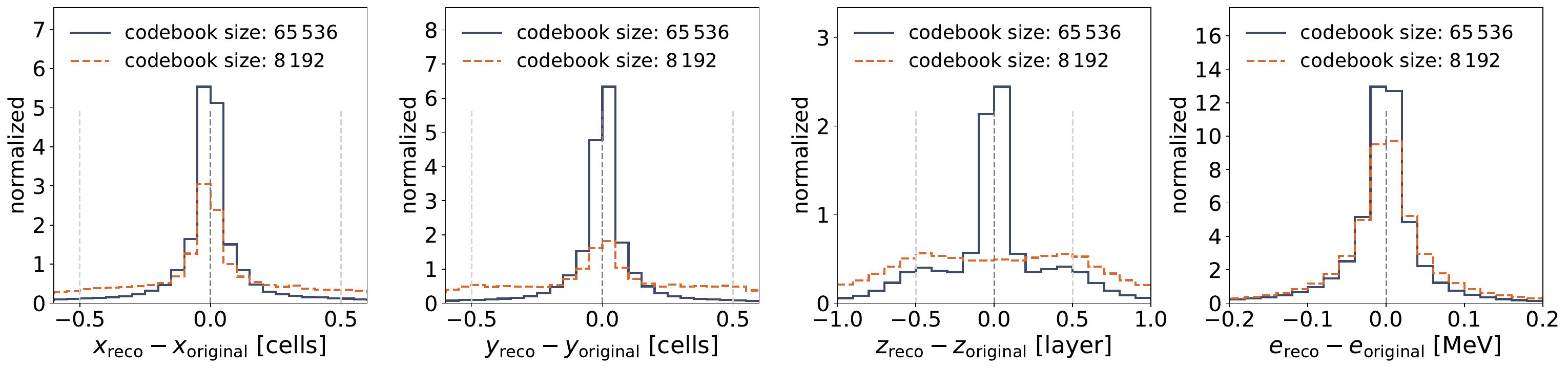}
   \caption{Reconstruction resolution for the input features ($x$, $y$, $z$, energy) for different codebook sizes.}
   \label{fig:reconstruction}
\end{figure}

We first investigate the encoding and decoding capabilities of the VQ-VAE.
To judge the effect of the tokenization and potential loss of information, we
compare the individual hits in the original showers with the corresponding hits in the reconstructed showers. A
perfect reconstruction would yield a Dirac delta function for the difference
between reconstructed and original values for each feature. However, as
shown in Figure~\ref{fig:reconstruction}, while the distributions surrounding the
center are indeed narrow, they do have some spread. A codebook size of 65\,536 shows a narrower resolution
distribution than a codebook size of 8\,192. In particular, the reconstruction of
$z$ for the latter has a larger spread of $\sigma_{\text{8\,192}}^{\text{$z$}} =
0.66$~layers compared to $\sigma_{\text{65\,536}}^{\text{$z$}} = 0.4$~layers with
the larger codebook size. For the energy, the respective spread values are
$\sigma_{\text{8\,192}}^{\text{energy}} = 0.11$\,MeV and
$\sigma_{\text{65\,536}}^{\text{energy}} = 0.07$\,MeV. Furthermore, the
reconstructed~$z$ distribution demonstrates a broader spread and a more complex
reconstruction relative to the transverse coordinates $x$ and $y$, which exhibit
similar and narrower distributions. This difference in reconstruction accuracy
can be attributed to a broader spatial extent of the showers along the
longitudinal axis~$z$.
However, because voxels are discrete, the three spatial
features need to be rounded to integers. Perfect resolution is achieved if these values remain within $\pm0.5$ before rounding, the region indicated by the light gray lines
in Figure~\ref{fig:reconstruction}. 

\begin{figure}[htbp]
    \centering 
    \includegraphics[width=\textwidth]{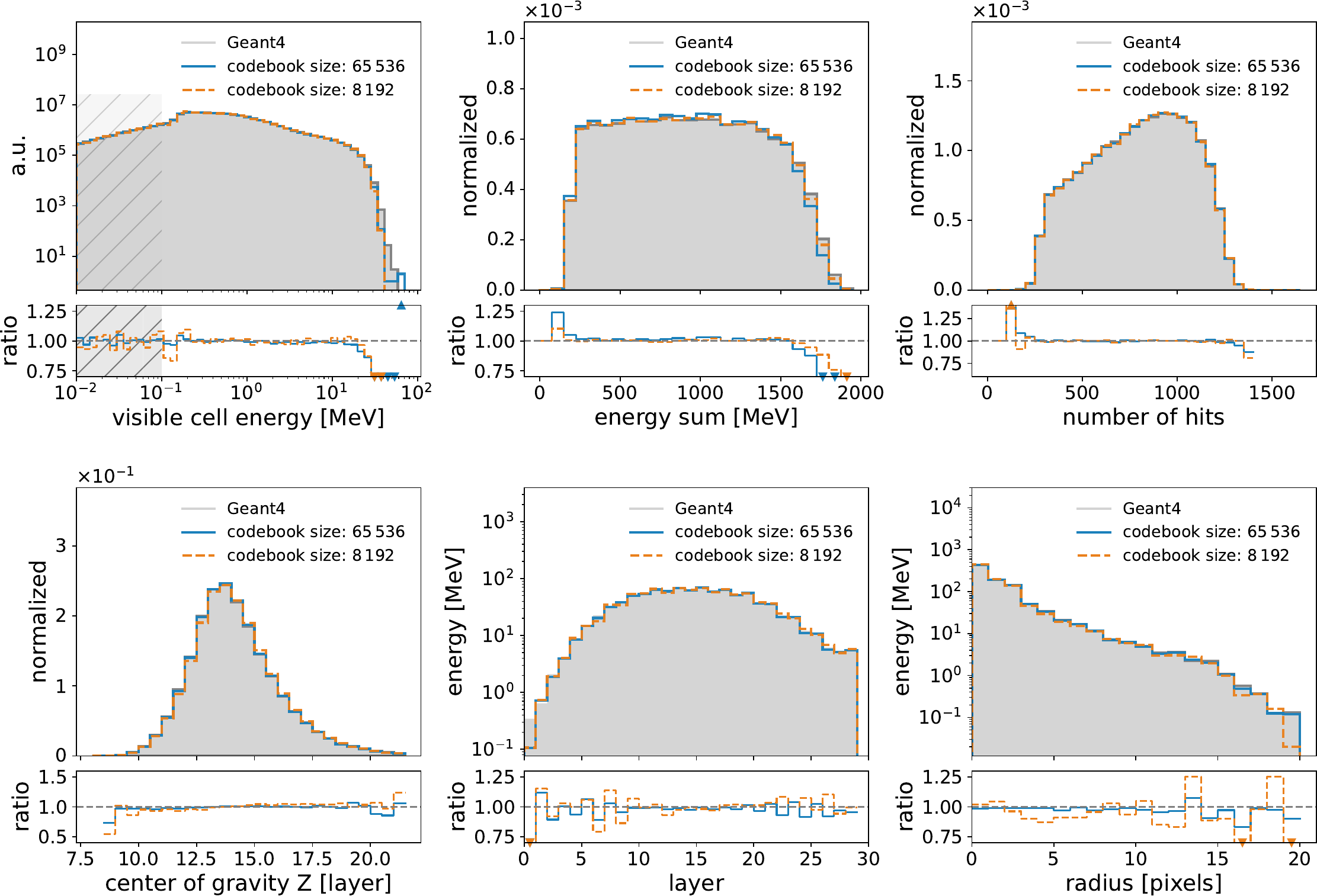}
    \caption{
        Distributions of physical observables between Geant4
        (grey, filled) with the codebook size of 65\,536 (blue) and codebook size
        of 8\,192 (orange). Hits that were below the MIP threshold ($0.1$\,MeV),
        i.e. those in the shaded region of the visible cell energy plot, were
        not considered for the comparison in the remaining distributions. This
        cutoff can affect the number of hits for reconstructed showers.
    }
    \label{fig:codebook_sizes}
\end{figure}

To accurately compare the reconstructed showers with the original showers,
we need to apply postprocessing. This step is explained in
Appendix~\ref{sec:appendix_postprocessing} and essentially projects hits back
into the voxel grid and processes duplicate hits (hits that are identical in all
of the three spatial features). For
the following analysis, showers are converted to tokens and then back to physical space.
Figure~\ref{fig:codebook_sizes} shows different feature distributions of the original and reconstructed showers, showcasing an overall good agreement between the two. Rare tokens, such as
those located at the edges of the shower or tokens associated with high-energy
hits, exhibit the lowest reconstruction quality. Again the VQ-VAE with the codebook size of 65\,536 performs better and has the smallest loss of information and is selected for tokenizing the showers for the generative training.

\subsection{Shower generation}
\label{sec:generation}
Following training, \textsc{OmniJet}-$\alpha_C$ generates point clouds autoregressively. Initialized
with a start token (a special token that initiates the autoregressive
generation process), the model predicts the probability distribution for the
next token based on the preceding sequence. \textsc{OmniJet}-$\alpha_C$ then samples from this
distribution, appending the chosen token to the growing sequence. This process
continues until a stop token (a special token that represents the end of the
generated sequence) is generated or the maximum sequence length of 1700 tokens
is reached. 
Unlike most ML-based shower generators, \textsc{OmniJet}-$\alpha_C$ is not trained to generate showers
for specific incident photon energies. Instead, the model learns to generate
showers with a variety of energies. We reserve a study of how to condition the
model on the incident energy for future work. This would allow the user to
request showers of a specific energy. In this first version however, we will
only compare the full spectrum of showers.

\begin{figure}[htbp]
   \centering 
   \includegraphics[width=\textwidth]{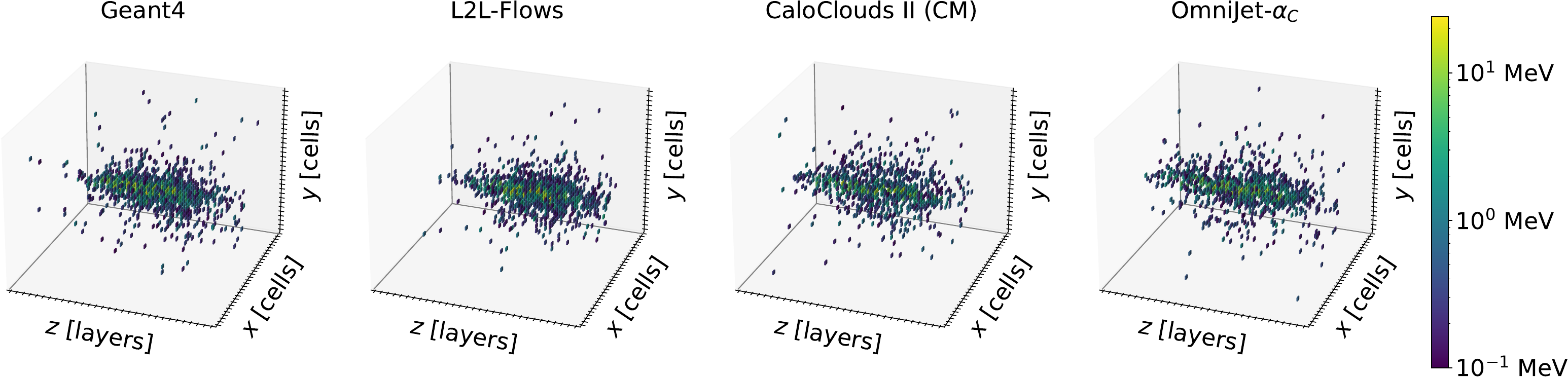}
   \caption{
        Examples of individual photon showers with a total energy sum of
        $1000$\,MeV generated by Geant4 (left), \textsc{L2LFlows} (center left), the
        \textsc{CaloClouds~II} (CM) (center right) and \textsc{OmniJet}-$\alpha_C$ (right).
    }
   \label{fig:3d_shower}
\end{figure}

We see in Figure~\ref{fig:3d_shower} that \textsc{OmniJet}-$\alpha_C$, \textsc{CaloClouds~II} (CM) and \textsc{L2LFlows} generate showers
that appear to be visually acceptable compared to Geant4. Next, we compare the
performance of \textsc{OmniJet}-$\alpha_C$ to \textsc{CaloClouds~II} (CM) and \textsc{L2LFlows} for three
different quantities\footnote{Note that compared to the original training of \textsc{CaloClouds~II} in Ref.~\cite{Buhmann:2023kdg}, this training is done at physical, ie. lower, resolution.}.

\begin{figure}[htbp]
    \centering 
    \includegraphics[width=\textwidth]{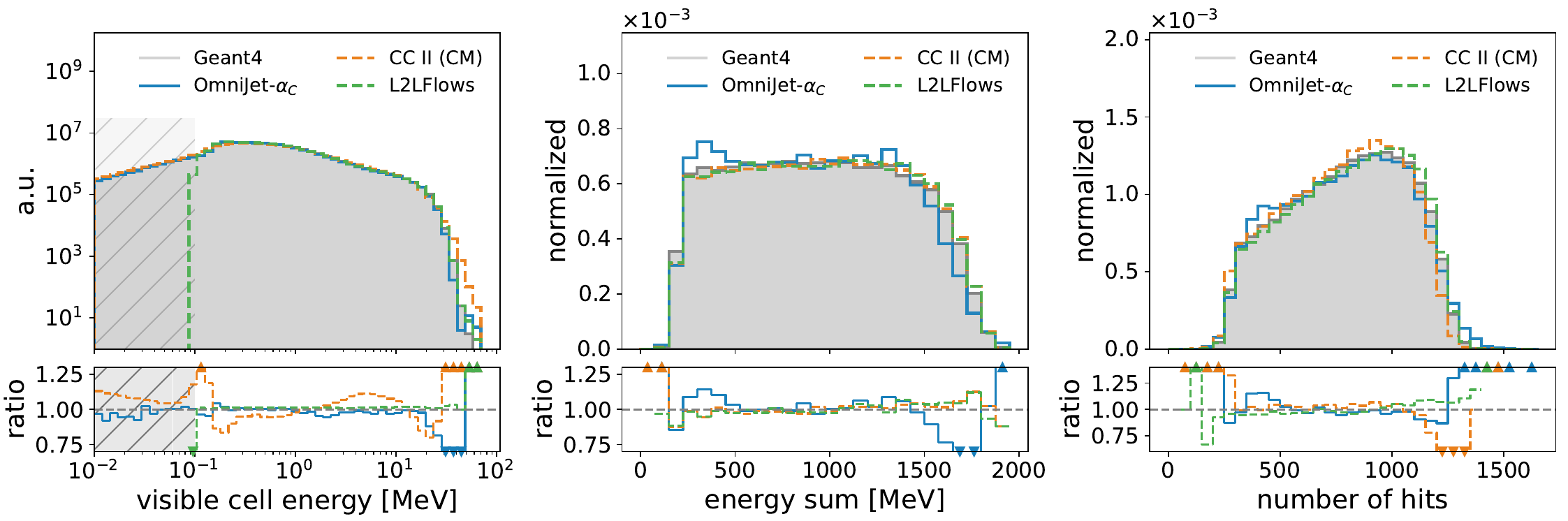}
    \caption{
        Distributions of  per-cell energy (left), total energy sum
        (middle) and the number of hits above $0.1$\,MeV (right)  between Geant4
        (grey, filled) and the generative models: \textsc{OmniJet}-$\alpha_C$ (blue), the \textsc{CaloClouds~II} (CM)
        (orange, dashed) and \textsc{L2LFlows} (green, dashed).
    }
    \label{fig:compare_gen_showers_top}
\end{figure}

Figure~\ref{fig:compare_gen_showers_top} (left) compares the 
cell energies. We
observe an accurate performance of \textsc{OmniJet}-$\alpha_C$ across almost the entire energy range, 
on par with \textsc{L2LFlows}. For the higher energies we see some deviations for both \textsc{OmniJet}-$\alpha_C$ and \textsc{CaloClouds~II} (CM). As seen in Figure~\ref{fig:codebook_sizes}, the mismodeling for \textsc{OmniJet}-$\alpha_C$ is introduced by the VQ-VAE. The behavior of \textsc{CaloClouds~II} (CM) is consistent with what was seen in the original paper. The shaded
area in the histogram corresponds to the region below half the energy of a minimal ionizing
particle (MIP). In real detectors, read-outs at such small energies are
dominated by noise. 
Therefore, cell energies below $0.1$\,MeV will not be considered in the
following discussion, and the remaining plots and distributions only include cells above this cut-off.

Figure~\ref{fig:compare_gen_showers_top} (center) shows the distribution of the
total energy sum of showers. For this calculation, the energy of all hits
surpassing half the MIP energy are added up for each shower. This distribution
is strongly correlated to the incident photon energy on which \textsc{L2LFlows} and \textsc{CaloClouds~II} (CM)
are conditioned. \textsc{OmniJet}-$\alpha_C$ has to learn this distribution on its
own.

Finally, Figure~\ref{fig:compare_gen_showers_top} (right) shows the number of hits. 
While the \textsc{L2LFlows} and \textsc{CaloClouds~II} (CM) are conditioned on this distribution, \textsc{OmniJet}-$\alpha_C$
is able to achieve good agreement with the Geant4 distribution without this conditioning. The discrepancies we see are a small
peak at a shower length of around 400 to 500, and also some showers that are too long.

\begin{figure}[htbp]
    \centering 
    \includegraphics[width=\textwidth]{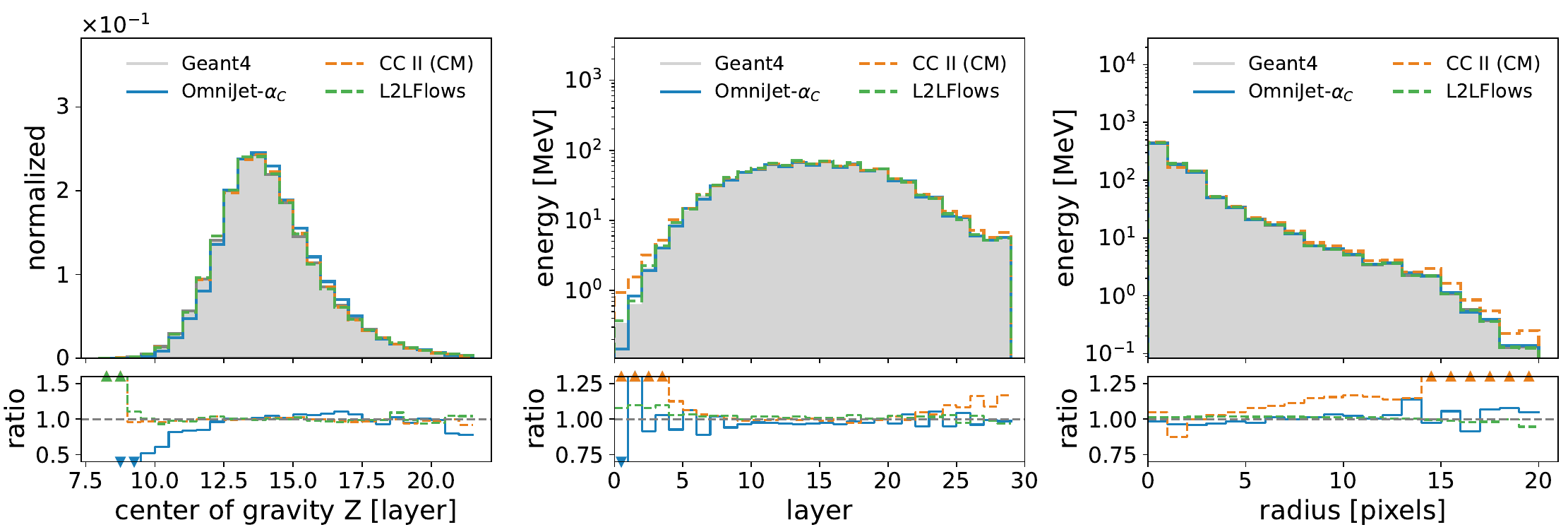}
    \caption{
        Distributions of center of gravity (left), mean energy per
        layer (middle) and the mean energy per layer between Geant4
        (grey, filled) and the generative models: \textsc{OmniJet}-$\alpha_C$ (blue), the \textsc{CaloClouds~II} (CM) (orange, dashed) and \textsc{L2LFlows} (green, dashed).
    }
    \label{fig:compare_gen_showers_bottom}
\end{figure}

In Figure~\ref{fig:compare_gen_showers_bottom} we compare the spatial properties
of the shower. The left plot shows that the Geant4 distribution of the
center of gravity along the \mbox{$z$-axis} is well modeled by all three architectures.
\textsc{OmniJet}-$\alpha_C$ performs better in the center of the
peak than at the edges.

The longitudinal energy distribution, depicted in the middle plot of
Figure~\ref{fig:compare_gen_showers_bottom}, reveals a comparatively weaker
performance of the \textsc{OmniJet}-$\alpha_C$ model and \textsc{CaloClouds~II} (CM) compared to L2LFlows in the initial 10 layers. 
However, \textsc{OmniJet}-$\alpha_C$ outperforms \textsc{CaloClouds~II} (CM) in the first 4 layers.
The mismodeling of \textsc{OmniJet}-$\alpha_C$ in the initial layers is likely attributable to the tokenization process (see
Figure~\ref{fig:codebook_sizes}), where these layers, being less common, are
represented by a limited number of tokens. A similar degradation is observed in the outer regions of the radial energy distribution (right plot
of Figure~\ref{fig:compare_gen_showers_bottom}), although \textsc{OmniJet}-$\alpha_C$ still outperforms \textsc{CaloClouds~II} (CM). %

Another important aspect for comparing generative models is the single-shower
generation time. Generating 1000 showers with \textsc{OmniJet}-$\alpha_C$, randomly sampled across all incident
energies, resulted in a mean and standard deviation of \SI{2.9295 \pm
1.0356}{\second} per shower. The generation was performed with a batch size of 2
on an NVIDIA\textsuperscript{\textregistered} A100 GPU. In contrast, Geant4 on a
CPU required \SI{4.08 \pm 0.17}{\second} per shower \cite{Buhmann:2020pmy}.
Therefore, our model demonstrates a speedup factor of $1.39$ in this case. On
identical hardware and with a batch size of 1000, L2LFlows achieves per-shower
generation times of  \SI{3.24 \pm 0.05}{\milli\second} and a speedup factor of
$1260$.
\textsc{CaloClouds~II} on identical hardware but with a batch size of 100 generates one
shower in \SI{16 \pm 6}{\milli\second} and achieves a speedup factor of $255$.
The comparatively slow performance of \textsc{OmniJet}-$\alpha_C$ is attributable to the characteristic quadratic scaling $\mathcal{O}(N^2)$ of the autoregressive transformer architecture with respect to the sequence length. Since this study did not prioritize
generation speed, optimizations such as multi-token generation are left for
future work.

\section{Conclusion}
\label{sec:conclusions}

In this work, we take a first important step towards building a foundation model
for several subdomains of particle physics. We show that we are able to use the
architecture and workflow of a foundation model originally developed for jet
physics to generate electromagnetic showers in a calorimeter, a fundamentally
different problem. This is a notable difference to previous efforts for
foundation models in HEP, which so far focused on tasks within one subdomain,
mostly different tasks within jet physics. Our work demonstrates that the same architecture can be successfully reused across distinct physical domains, without significant modifications. It is also the first implementation of
a GPT-style autoregressive generative model for calorimeter shower point cloud generation.

The next immediate step will be to investigate whether this model can be used
for transfer learning between different types of showers. In the long term, we
aim to develop a joint model that can work with both jets and showers. Combining
tasks from different subdomains in one single framework is a necessary step
towards a foundation model for particle physics that can handle a variety of
data types and tasks.

\section*{Acknowledgements}
The authors would like to thank William Korcari for support with the dataset, as well as Thorsten Buss for providing the \textsc{L2LFlows} samples. JB, AH, GK, MM and HR are supported by the DFG under the German Excellence Initiative -- EXC 2121  Quantum Universe – 390833306, and by PUNCH4NFDI – project number 460248186. This work has used the Maxwell computational resources at Deutsches Elektronen-Synchrotron DESY, Hamburg, Germany.

\section*{Code Availability}
The code for this paper can be found at
\href{https://github.com/uhh-pd-ml/omnijet_alpha_c}{\texttt{https://github.com/uhh-pd-ml/omnijet\_alpha\_c}}.
The code for recreating the dataset can be found at \\
\href{https://github.com/FLC-QU-hep/getting_high}{\texttt{https://github.com/FLC-QU-hep/getting\_high}}.

\appendix
\section{Model details and hyperparameters}
\label{sec:appendix_hyperparams}

Different hyperparameter configurations were tested for the individual model
components of \textsc{OmniJet}-$\alpha_C$. The configurations presented in the following were found to
lead to stable trainings. However, no extensive hyperparameter optimization was
performed.
\begin{table}[htbp] %
    \centering
    \caption{Hyperparameters used in the VQ-VAE training.}
    \label{tab:hyperparametersVQ-VAE}
    \begin{tabular}{l r}
        \hline
        \textbf{Hyperparameter} & \textbf{Value} \\ \hline
        Learning rate & 0.001 \\ 
        Optimizer & Ranger \\ 
        Batch size & 152 \\ 
        Batches per epoch & 1000 \\ 
        Number of epochs & 588 \\ 
        Hidden dimension & 128 \\ 
        Codebook size & 65\,536 \\ 
        $\beta$ & 0.8 \\
        $\alpha$ & 10 \\
        Replacement frequency & 100 \\
        \hline
    \end{tabular}
\end{table}

The hyperparameters used for the VQ-VAE training are shown in
Table~\ref{tab:hyperparametersVQ-VAE}. Only the
codebook size, replacement frequency and the hyperparameter $\beta$ were adjusted. The
remaining hyperparameters are the same as in \textsc{OmniJet}-$\alpha$.
An increase of the codebook size from $8\,192$ to $65\,536$
was found to improve the reconstruction capabilities (i.e. the resolution of the
tokenized showers). The codebook utilization, i.e. the fraction of used tokens,
is also monitored during the training to ensure that the resulting codebook is
used completely. Unused tokens would drastically increase the number of parameters
of the generative model while not adding any potential improvements in the performance
of the generative model.
The current setup results in a codebook utilization of the final VQ-VAE model of $99.65\%$. 
The hyperparameter $\beta$ which defines the relative
importance of how much weight should be given to updating the encoder embeddings
$z_e$ towards the codebook vectors $z_q$ and vice versa, is decreased from $0.9$ 
to $0.8$. 
This leads to a higher emphasis on adapting the encoder to bring the embeddings $z_e$
closer to the codebook vectors $z_q$.
Furthermore, the optimization process employs a token replacement strategy based on usage
frequency. The chosen replacement frequency of 100 batches (instead of 10)
indicates that a token must be used at least once within the preceding 100
batches to avoid being replaced by a new token. 
We used the Lookahead optimizer \cite{zhang2019lookaheadoptimizerksteps} with RAdam as the inner optimizer \cite{yong2020gradientcentralizationnewoptimization}.

For the hyperparameters of the backbone, no changes compared to \textsc{OmniJet}-$\alpha$ were made
except for the batch size. The hyperparameters used are listed in
Table~\ref{tab:hyperparametersOJA}.

\begin{table}[htbp] %
    \centering
    \caption{Hyperparameters used in the generative model training.}
    \label{tab:hyperparametersOJA}
    \begin{tabular}{l r}
        \hline
        \textbf{Hyperparameter} & \textbf{Value} \\ \hline
        Learning rate & 0.001 \\ 
        Optimizer & Ranger \\ 
        Batch size & 72 \\ 
        Batches per epoch & 6000 \\ 
        Number of epochs & 106 \\ 
        Embedding dimension & 256 \\ 
        Number of heads & 8 \\ 
        Number of GPT blocks & 3 \\ 
        \hline
    \end{tabular}
\end{table}

\section{Postprocessing}
\label{sec:appendix_postprocessing}
\begin{figure}
    \centering 
    \includegraphics[width=0.8\textwidth]{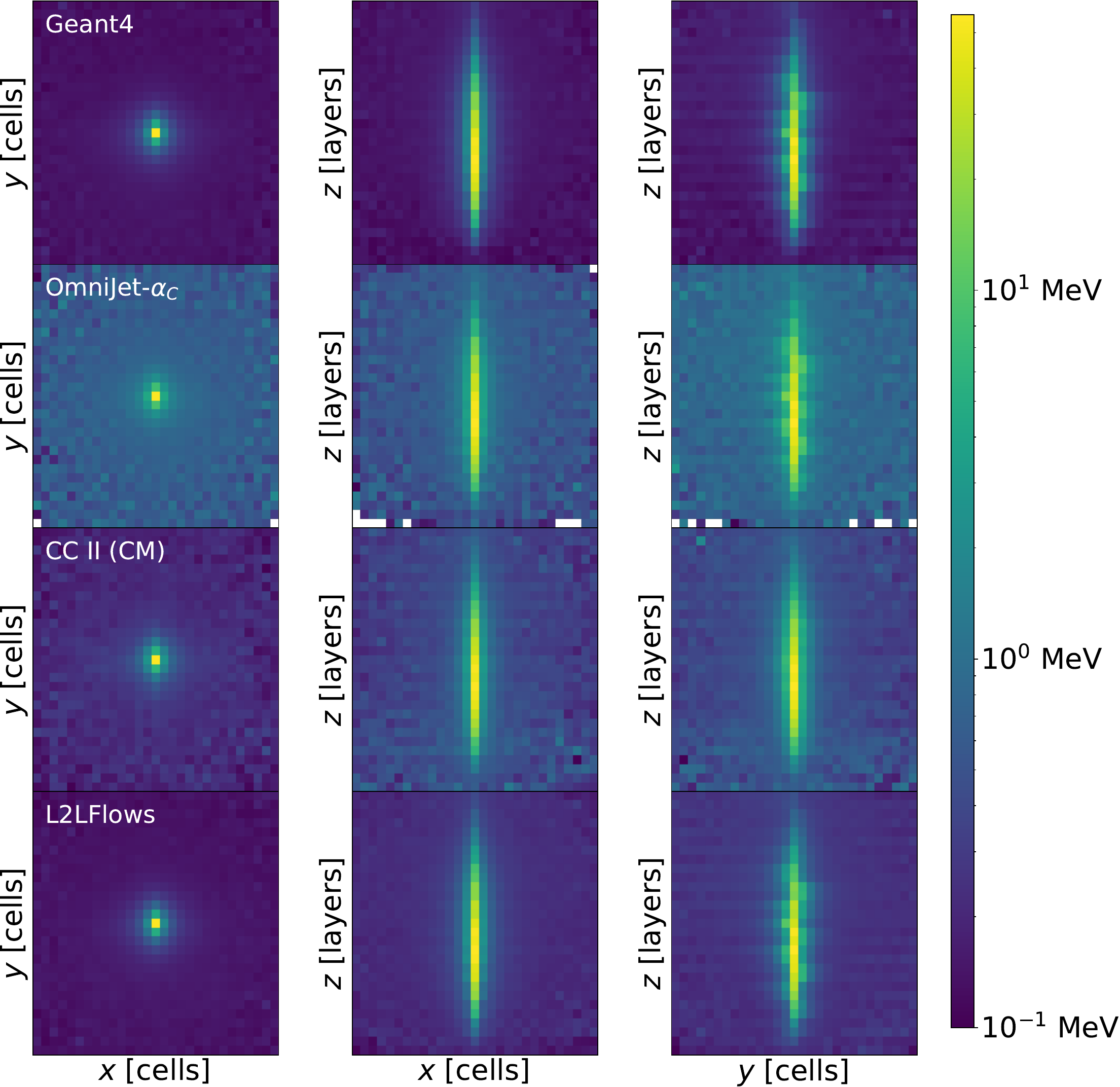}
    \caption{
        Overlay of 10k showers for all simulators for the full spectrum, where the
        voxel energies are summed along the $z$- (left), $y$- (middle) and $x$-axis
        (right). In all plots, the mean over the number of showers is taken. 
    }
    \label{fig:resolution}
\end{figure}

Projecting the hits of a point cloud model back onto the voxel grid can result in duplicate hits in some voxels. To resolve these duplicates, the voxels with lower
energy are translated along the $z$-axis to the nearest unoccupied voxel
position. This heuristic preserves both the total energy and the hit count while
minimally impacting the $z$-distribution. We could also translate the voxels
along the $x$- or $y$-axis, but as shown in Figure~\ref{fig:resolution} the hit
energies are not invariant in these directions.

\section{Generation quality}
\label{sec:generation_quality}
\subsection{Generation}
\label{sec:Corelation}
\begin{figure}[h]
    \centering 
    \includegraphics[width=\textwidth]{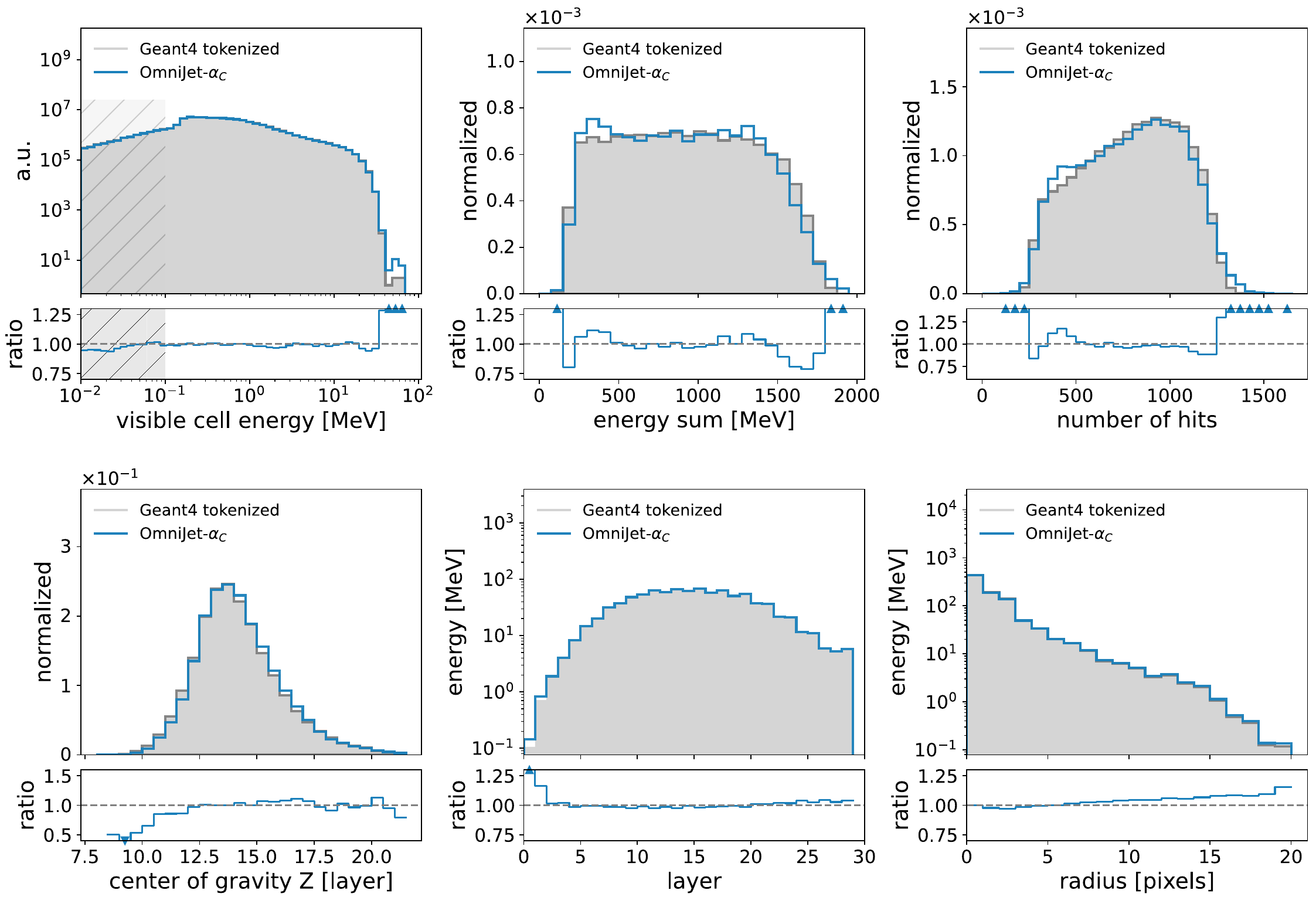}
    \caption{
        Distributions for Geant4 after tokenization and reconstruction (gray) and generated showers by \textsc{OmniJet}-$\alpha_C$ (blue).
    }
    \label{fig:compare_gen_showers_token}
\end{figure}

To isolate the impact of the tokenization and subsequent reconstruction from that of the generative model itself, Figure~\ref{fig:compare_gen_showers_token} compares showers generated by \textsc{OmniJet}–$\alpha_C$ to Geant4 showers that were first tokenized and then reconstructed. This figure should be viewed in context of Figure~\ref{fig:codebook_sizes}, which shows the performance of the tokenization and reconstruction, and Figures~\ref{fig:compare_gen_showers_top} and \ref{fig:compare_gen_showers_bottom} which show the generated showers compared to the original Geant4 showers. 
The same discrepancy in the energy sum and number of hits that was seen in Figure~\ref{fig:compare_gen_showers_top} can also be seen here. This means that this is an effect of the generative model itself, not the VQ-VAE.
\subsection{Correlations}
\label{sec:Corelation}
\begin{figure}[htbp]
    \centering 
    \includegraphics[width=0.85\textwidth]{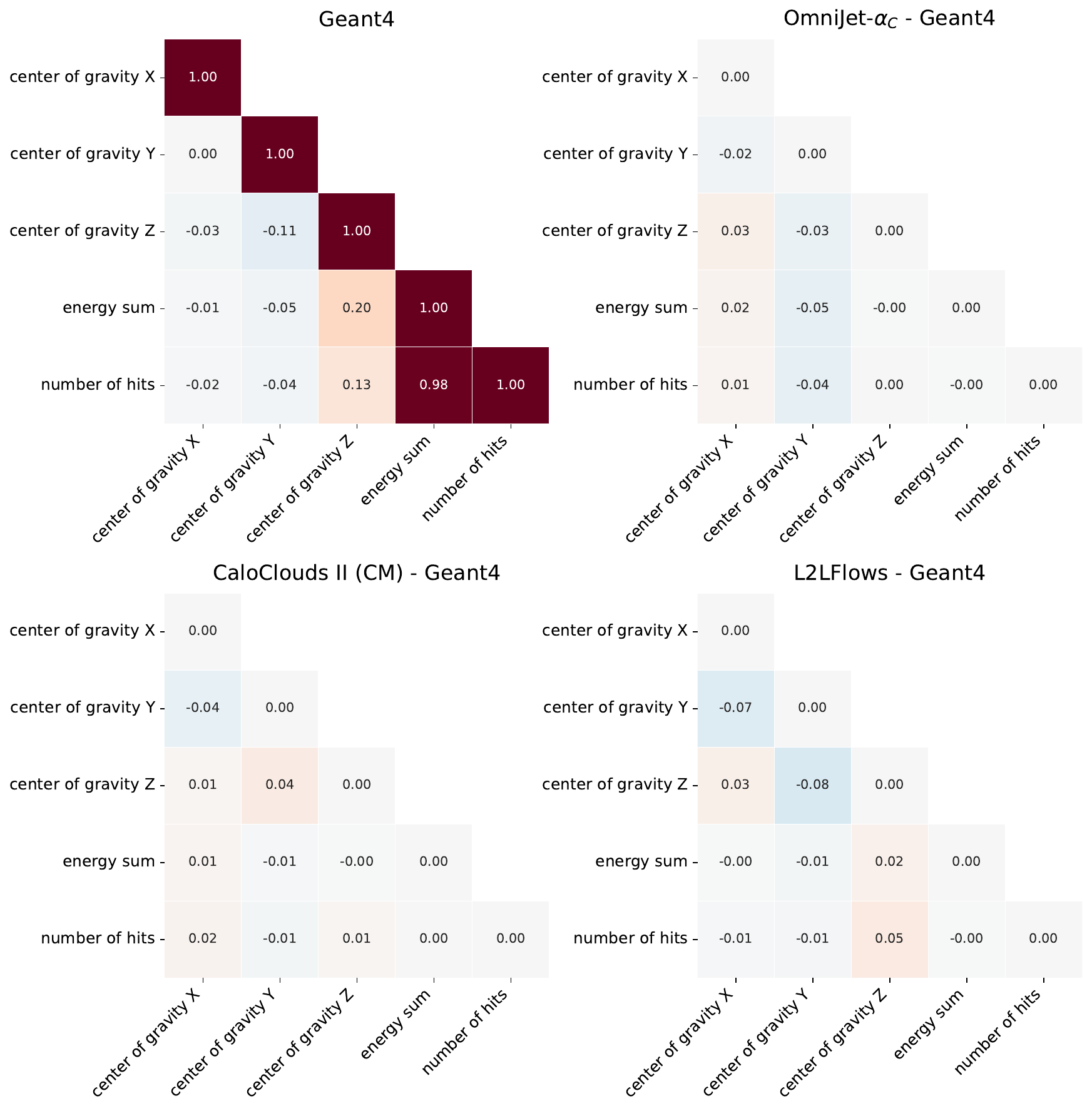}
    \caption{
        Linear correlation of shower features for Geant4 (top left) and the residuals for \textsc{OmniJet}-$\alpha_C$ (top right), \textsc{CaloClouds~II} (CM) (bottom left) and \textsc{L2LFlows} (bottom right).
    }
    \label{fig:correlations}
\end{figure}

Figure~\ref{fig:correlations} summarizes how well each model preserves the linear relationships among shower observables. The relation between number of hits and energy sum, which is an important relation in the calorimeter context, is preserved in all models. The correlations between the other shower features are weak in the Geant4 baseline, a characteristic that is also reflected by all models.

\newpage

\bibliographystyle{JHEP}
\bibliography{biblio.bib}

\end{document}